\documentclass[useAMS,usegraphicx,usenatbib]{mn2e}
\usepackage{aas_macros}     
\usepackage{amsmath}

\usepackage{fixltx2e}[1999/12/01]

\newcommand{\Mass}{ \ensuremath{ h^{-1} M_{\odot}} }
\newcommand{\Mpc}{ \ensuremath{h^{-1} {\rm Mpc}} }
\newcommand{\Gpc}{ \ensuremath{h^{-1} {\rm Gpc}} }
\newcommand{\hMpc}{ \ensuremath{ h {\rm Mpc^{-1}}} }
\newcommand{\kpc}{ \ensuremath{ h^{-1} {\rm kpc}} }
\newcommand{\Rnl}{ R_{\rm nl}}
\newcommand{\zf}{ z_{\rm f}} 
\newcommand{\mps}{ m_{\rm p}} 

\newcommand{\vect}[1]{\bmath{#1}}

\title[One simulation to fit them all]
{One simulation to fit them all - changing the background parameters of a cosmological N-body simulation}
\author[Angulo \& White]{
\parbox[h]{160mm}{
R. E. Angulo\thanks{reangulo@mpa-garching.mpg.de} \& S. D. M. White
}\vspace{6pt}\\
Max Planck Intitute fur Astrophysik, D-85741 Garching, Germany.
\vspace*{-0.5cm}}

\begin{document}
\date{\today}
\pagerange{\pageref{firstpage}--\pageref{lastpage}} \pubyear{2009}
\maketitle
\label{firstpage}

\begin{abstract} 
We demonstrate that the output of a cosmological N-body simulation can, to
remarkable accuracy, be scaled to represent the growth of large-scale structure
in a cosmology with parameters similar to but different from those originally
assumed. Our algorithm involves three steps: a reassignment of length, mass and
velocity units, a relabelling of the time axis, and a rescaling of the
amplitudes of individual large-scale fluctuation modes. We test it using two
matched pairs of simulations. Within each pair, one simulation assumes
parameters consistent with analyses of the first-year WMAP data. The other has
lower matter and baryon densities and a 15\% lower fluctuation amplitude,
consistent with analyses of the three-year WMAP data. The pairs differ by a
factor of a thousand in mass resolution, enabling performance tests on both
linear and nonlinear scales. Our scaling reproduces the mass power spectra of
the target cosmology to better than 0.5\% on large scales ($k < 0.1 h{\rm
Mpc}^{-1}$) both in real and in redshift space. In particular, the BAO features
of the original cosmology are removed and are correctly replaced by those of
the target cosmology. Errors are still below 3\% for $k < 1 h{\rm Mpc}^{-1}$.
Power spectra of the dark halo distribution are even more precisely reproduced,
with errors below 1\% on all scales tested. A halo-by-halo comparison shows
that centre-of-mass positions and velocities are reproduced to better than
$90h^{-1}$kpc and 5\%, respectively. Halo masses, concentrations and spins
are also reproduced at about the 10\% level, although with small biases. Halo
assembly histories are accurately reproduced, leading to central galaxy
magnitudes with errors of about $0.25$ magnitudes and a bias of about $0.13$
magnitudes for a representative semi-analytic model. This algorithm will enable
a systematic exploration of the coupling between cosmological parameter
estimates and uncertainties in galaxy formation in future large-scale structure
surveys.
\end{abstract}

\section{Introduction}

Ever since the first measurements, made almost 40 years ago, the clustering of
galaxies has been recognized as a powerful and robust tool to investigate some
of the most fundamental questions in cosmology and galaxy formation.
Subsequent years have produced major advances on both observational and
theoretical fronts, and the study of the three-dimensional distribution of
galaxies has contributed enormously to our current understanding of cosmic
evolution. For instance, it has provided information about the total matter
content of the Universe, the amount of baryons, the formation and evolution of
galaxies, and even about the nature of gravity.

The potential to learn from the galaxy distribution is still far from
exhausted.  New galaxy surveys are scanning ever larger volumes providing more
homogeneous and more detailed structural information, and thereby improving
considerably the accuracy which can be achieved with existing probes of
cosmological physics. In addition, exciting new probes such as the imprint of
Baryonic Acoustic Oscillations \citep[BAO, e.g.][]{Eisenstein2005, Cole2005,
Gaztanaga2008, Sanchez2009}, the form of redshift-space distortions \citep{Guzzo2008,
Percival2008}, the clustering amplitude of dark matter haloes
\citep{Seljak2008, Matarrese2008, Carbone2008} and the evolution of
gravitational lensing signals \citep{Huterer2006, Hoekstra2006, Berge2009,
Debono2009} have been proposed to explore, for instance, the nature of the dark
energy and the amplitude of primordial non-Gaussianities. In addition,
higher-order clustering could provide fresh insights into many of the
cosmological issues mentioned above \citep{Baugh2004, Croton2007, Ross2008}.

A key step towards achieving optimal exploitation of galaxy clustering data and
towards assessing the power of future experiments is to refine our
understanding of the connection between galaxy formation and the background
cosmological model.  In particular, there are two aspects that need to be
addressed. The first is the interplay between local baryonic physics and the
large-scale evolution of density perturbations \citep[see e.g.][]{Zheng2007,
Harker2007}. For instance, how different is the efficiency of star formation in
a universe with larger dark matter perturbation amplitudes? Can we use the
small-scale clustering of galaxies, which is heavily affected by astrophysics,
to learn about the underlying cosmological model?  Another crucial aspect is to
study whether uncertainties in galaxy formation physics will weaken or possibly
bias the cosmological constraints extracted from future datasets, e.g.  how
robust is a measurement of the dark energy equation of state derived from BAO
to uncertainties in how (and where) galaxies form \citep[e.g.][]{Angulo2008}?

In order to explore these issues we require accurate and flexible theoretical
predictions for the properties and the spatial distribution of galaxies over
large volumes. Currently, the most sophisticated models are built in a
three-step process. First, the abundance, structure and clustering of haloes
are predicted by following the evolution of the dark matter in a
high-resolution N-body simulation \citep[e.g.][]{Springel2006}.  Second, the
properties of galaxies within these haloes are predicted using either
semi-analytic recipes \citep[e.g.][]{Baugh2006} or empirically motivated
parametric techniques known as halo occupation distribution (HOD) models
\citep[e.g.][]{Cooray2002}. Lastly, the appropriate flux limits, sample
selection criteria, redshift completeness and survey geometry are applied to
build realistic mock catalogues. 

Unfortunately, the realism of these mock catalogue implies a substantial
computational cost, especially when very large scales are probed.  A simulation
that resolves simultaneously Gigaparsec scales and Milky-Way sized haloes
takes at least several days in the largest supercomputers available today.
The modelling becomes even more expensive when one considers that a large
number of independent N-body simulations are needed to assess clustering errors
properly. An adequate estimation of the variance requires several dozen
realizations of the density field (e.g. a $10$\% error on a variance estimate
for a Gaussian distribution requires $\sim 50$ realizations), and an order of
magnitude more simulations are needed to robustly compute the full covariance
matrix \citep{Takahashi2009}. In practice, these technical constraints mean
that this method can be applied to simulate galaxies only in a very limited
number of cosmological scenarios, which then very sparsely sample the allowed
cosmological parameter space. This results in an incomplete picture of the
coupling between galaxy formation and cosmology.

In this paper we propose and test a new approach which avoids this limitation
and allows high-fidelity mock catalogues to be created for a broad range of
cosmological scenarios.  Current constraints on cosmological parameters,
derived from analyses of CMB fluctuations, of supernovae and of galaxy
clustering, allow only relatively small variations  around the preferred
structure formation model \citep[see e.g][]{Sanchez2009}. Taking advantage of
this, we explore whether it is possible to manipulate the results of a {\it
single} N-body simulation to represent the growth of structure (with the
accuracy needed for faithful galaxy modelling) over the {\it whole} allowed
cosmological parameter space. In particular, we will focus on reproducing the
position, velocity, abundance and mass of the dark matter haloes thought to
host galaxies.  Matching these properties accurately at all times ensures a
correct merger history for all DM haloes and therefore the correct prediction
of the properties of the galaxies within them for any given galaxy formation
model. The resulting ``scaled" galaxy catalogues thus correspond to those
expected in the target cosmology.
 
Our proposed algorithm consists of the following steps. First, a scaling of the
box-size and particle mass together with a reassignment of the output redshifts
of the original N-body simulation matches the shape of the linear variance of
fluctuations in the target cosmology over the range of scales which give rise
to observable {\it nonlinear} structure in that cosmology.  Second, using
linear perturbation theory, the large-scale components of the displacement and
velocity fields are corrected to take account of the differences in power
spectrum shape between the original and target cosmologies. 

The layout of our paper is as follows. In \S2 we present more details of our
algorithm. We discuss how length, mass and velocity units are altered and the
time axis relabelled to match the nonlinear evolution of perturbations.  We
then describe how to rescale the amplitude of individual long-wavelength modes
at each redshift in order to correct the linear and quasi-linear evolution on
these scales. In \S3.1 we present a suite of N-body simulations specially
designed to test the accuracy of our scaling procedure and in \S3.2 we use them
to show how well the full nonlinear matter power spectrum and its evolution can
be reproduced for a chosen cosmology by scaling the particle distribution from
a simulation of a different cosmology.  The resulting halo catalogues are then
compared in \S4 with catalogues constructed from a matched simulation carried
out directly in the target cosmology.  Finally, in \S5 we provide a summary and
we discuss applications of our method.  

\section{Scaling an N-body simulation}

In this section we will show how it is possible to use the results of an N-body
simulation carried out in one cosmology to represent the growth of structure in
a different cosmology. For example, how to predict the evolution of structure in a
cosmology with the parameters preferred by analyses of the 3rd year WMAP data
using the Millennium Simulation, which was carried out using parameters based
on an analysis of the 1st year WMAP data. This procedure can be repeated for an
arbitrary number of target cosmologies, fully covering the allowed region of
the cosmological parameter space for the standard $\Lambda$CDM model.  In \S2.1
we provide the background and motivation behind our algorithm, and in \S2.2 we
outline a practical implementation of these ideas.

\subsection{Theoretical motivation}

Let us denote by $P(k)$ the linear matter power spectrum extrapolated to $z=0$
in our original cosmology (WMAP1 in the example we give below). The linear
theory power spectrum at redshift $z$ is then $D^2(z)P(k)$, where $D(z)$ is the
linear growth factor for this original cosmology in units of its present-day
value, i.e. $D(0)=1$.  We can now define the variance of the linear density
field as a function of smoothing scale through:

\begin{equation}
\sigma^2(R,z) \equiv 
	D^2\frac{1}{4\pi} \int_{0}^{\infty} {\rm d}k k^2 P(k) W^2(kR) = D^2 \sigma^2_0(R)  
\end{equation}

\noindent where $R$ is a comoving smoothing scale, and $W(kR) = 3 (\sin(k R)
-k R \cos(k R))/(kR)^3$ is the Fourier transform of the real-space top-hat
filter. 

For a specific realisation of $P(k)$, the approximation of \cite{Zeldovich1970}
relates Eulerian comoving positions $\vect{x}$ at redshift $z$ to the
corresponding initial, or Lagrangian, comoving positions $\vect{q}$ through:

\begin{equation}
\vect{x}(\vect{q},z) = \vect{q} - D(z)\vect{S}(\vect{q}),
\label{eq:zeldovich_x}
\end{equation} 

\noindent and in this approximation, the peculiar velocities at redshift $z$ are 
given by:

\begin{equation}
a \dot{\vect{x}} \equiv \vect{v}(\vect{q},z) =  - \frac{\dot{D}(z)}{1+z} \vect{S}(\vect{q}).
\label{eq:zeldovich_v}
\end{equation} 

\noindent The dots here denote derivatives with respect to time and
$\vect{S}(\vect{q})$ is the linear displacement field of the realisation
extrapolated to $z=0$.

Let us now consider a ``target" cosmological model (WMAP3 in our example below)
which we seek to study using simulation results obtained within our original
cosmological model. In the following, unprimed quantities will continue to
refer to the original cosmology, while primed quantities $P'$, $D'$, $\sigma'$,
$\sigma'_0$, $S'$ will refer to the target cosmology. Let us assume $\zf$ to be the
latest time at which the target cosmology must be matched (often $\zf=0$). The
first step in our procedure is to decide a length scaling from original
simulation box-size to target box-size $L' = s\,L$. This is defined so that the
linear fluctuation amplitude $\sigma'(R, \zf)$ in the target cosmology over the
range $[R_1, R_2]$ is as close as possible to that in the original cosmology
over the range $[s^{-1}\,R_1,s^{-1}\,R_2]$ at a redshift $z_*(\zf)$. Thus we
minimize:

\begin{eqnarray}
\delta_{\tt rms}^2 (s, z_*) &=& \frac{1}{\ln (R_2/R_1)} \times \nonumber\\ & &  
\int_{R_1}^{R_2} \frac{{\rm d}R}{R} \left[1 - \dfrac{\sigma(s^{-1}\,R,z_*)}{\sigma'(R,\zf)} \right]^2,
\label{eq:rms}
\end{eqnarray}

\noindent over $s$ and $z_*$ \citep[see][for a similar
approach]{Zheng2002,Harker2007}.

In Press-Schechter theory the halo mass function is determined completely by
the variation with scale of $\sigma(R,z)$ the smoothed linear variance of the
field.  Eq.\ref{eq:rms} is thus equivalent to minimizing the difference between
the halo mass functions in the target and scaled original cosmologies over the
(target) mass range $[M(R_2),M(R_1)]$. This sets the range of scales included
in Eq.\ref{eq:rms}: $M(s^{-1}R_2)$ is the mass of the largest halo in the
original simulation at $z=z_*$ and $M(s^{-1}R_1)$ that of the smallest resolved
halo. For the top-hat filter we are using we have
$M(R)=\frac{4\pi}{3}\rho_0\,R^3$.

We take the original simulation at $z_*$ after scaling all lengths by $s$,
to be our first approximation to a simulation of the target cosmology at $\zf$.
Earlier redshifts in the target cosmology ($z > \zf$) are then matched to those in
the original cosmology ($z > z_*$) through:

\begin{equation}
D'(z') = \frac{D(z)}{D(z_*)} D'(\zf).
\label{eq:gfact}
\end{equation}

If needed, peculiar velocities in the target cosmology can be obtained
approximately from those in the original cosmology through:

\begin{equation}
\vect{v} \rightarrow \vect{v'} = s \frac{\dot{D'}(z')}{\dot{D}(z)} \frac{(1+z)}{(1+z')} \vect{v},
\label{eq:linvel}
\end{equation}

\noindent but in practice we use a more accurate procedure which we now describe.

At each redshift in the target cosmology we can define $\Rnl(z')$ through
$\sigma'(\Rnl,z')=1$ or $\sigma_0'(\Rnl) = 1/D'(z')$.  The fluctuations in the
target cosmology on all scales larger than $\Rnl$ (i.e.  $k<\Rnl^{-1}$) are in
the linear regime and thus the corresponding displacements and velocities can
be corrected for residual differences in power spectrum between the two
cosmologies. This is done by subtracting the long wavelength components of the
particle position and velocity fields using the Ze'dovich approximation
(Eqs.~\ref{eq:zeldovich_x} and \ref{eq:zeldovich_v}) with the following
low-pass-filtered displacement field,  

\begin{equation}
D(z) \vect{S}_{\vect{k}}(\vect{q};s^{-1}\Rnl) = 
\begin{cases}
D(z) \vect{S}_{\vect{k}}(\vect{q}) & \text{for $|k| < s/\Rnl$},\\
					   0 & \text{for $|k| > s/\Rnl$},
\end{cases}
\end{equation}

\noindent and then adding them back in after scaling each Fourier component of
the displacement field by the square root of the ratio of the amplitude of
target to original power spectrum,

\begin{eqnarray}
D'(z') \vect{S}'_{\vect{k}}(\vect{q};\Rnl) = \nonumber \phantom{xxxxxxxxxxxxxxxxxxxxxxxxxxxxxx}\\
\begin{cases}
D(z) s \left[ \dfrac{P'(\vect{k})}{s^3 P(s\vect{k})} \right]^{1/2} \vect{S}_{s\vect{k}}(\vect{q}) & \text{for $ |k| < 1/\Rnl$},\\
					   0 & \text{for $ |k| > 1/\Rnl$},
\end{cases}
\label{eq:new_disp}
\end{eqnarray}

\noindent which is equivalent to the long-range displacement field in the
target cosmology. Note that the corresponding velocity fields are derived from
these displacement fields through Eq.~\ref{eq:zeldovich_v}.  Also note that, in
Fourier space, the low-pass-filtered displacement and velocity fields of the
original simulation at $z$ can be calculated from those used to generate the
initial conditions.

A similar technique has been previously studied in the literature, although for
a different purpose, by \cite{Tormen1996} and \cite{Cole1997}. These authors
used linear perturbation theory to add long-wavelength modes to an N-body
simulation of limited box-size. As noted by \cite{Cole1997}, a naive
implementation neglects the coupling of these long-wavelength modes to 
the smaller-scale perturbations already included in the simulation. This
produces systematic and large biases in the internal properties and clustering
of haloes, especially massive ones. We note that in our case this problem is
considerably alleviated since we impose relatively small changes on the
amplitudes of long-wavelength modes. The correct coupling between large and
small scales is included automatically to lowest order.










\subsection{Implementation}

We now describe briefly the steps of a practical implementation of the ideas
presented in the previous subsection. 

\begin{itemize}
\item[1)] Calculate the scale factor, $s$, and the redshift $z_*$ 
by minimizing Eq.~\ref{eq:rms}. 

\item[2)] For each output $z>z_*$ of the original simulation calculate the
corresponding $z'(z)$ in the target cosmology using Eq.~\ref{eq:gfact}.

Then for each $z'-z$ pair:

\item[3)] Calculate $\Rnl$ .
\item[4)] Calculate $\vect{S}(\vect{q};s^{-1}\Rnl)$ for the original simulation.
\item[5)] Invert $\vect{x} = \vect{q} - D(z)\vect{S}(\vect{q};s^{-1}\Rnl)$ in the
original simulation to obtain $\vect{q}(\vect{x};s^{-1}\Rnl, z)$. This can be done
recursively:

\begin{equation}
\vect{q}_i = \vect{x} - D(z) \vect{S}(\vect{q};s^{-1}\Rnl)
\label{eq:iter}
\end{equation}

\noindent with $\vect{q}_0 = \vect{x}$ as the starting condition and requiring
that $|\vect{q}_i-\vect{q}_{i-1}| < 0.01 \Mpc$, for example, for convergence.

\item[6)] Remove low-pass-filtered displacement and velocity fields from the
original simulation data at redshift $z$.
\item[7)] Scale to the target cosmology:
\item[7.1)] Box size: $L \rightarrow L' = sL$.
\item[7.2)] Redshift: $z \rightarrow z'$.
\item[7.3)] Particle mass: 

\begin{equation}
\mps \rightarrow \mps' = \frac{\Omega_{\rm m}' H'^2 L'^3}{ \Omega_{\rm m} H^2 L^3} \mps.
\end{equation}

\item[7.4)] Residual nonlinear velocities: 

\begin{eqnarray}
\vect{v} \rightarrow \vect{v}' &=& \left[ \frac{L}{ (1+z) \mps} \frac{(1+z') \mps'}{L'} \right]^{1/2} \vect{v},\\
                 &=& \left[ \frac{\Omega_{\rm m}' H'^2 L'^2 (1+z')}{ \Omega_{\rm m} H^2 L^2 (1+z)}  \right]^{1/2} \vect{v},
\end{eqnarray}

\noindent where unprimed and primed cosmological parameters are present-day values 
			in the original and target cosmologies, respectively.

\item[8)] Compute new long-range displacement and velocity fields by scaling
the Fourier components of the original fields according to
Eq.~\ref{eq:new_disp} using the power spectra of the two cosmologies. Using the
Zel'dovich approximation, add these low-pass filtered fields to the scaled
positions and velocities of the particles.  
\end{itemize}

\section{Numerical simulations}

In this section we provide details of the N-body simulations used to illustrate 
and probe the accuracy of our algorithm. First, we describe the numerical and
cosmological parameters of our simulations, which were carried out at two
different resolutions and in two different cosmologies.  Next, we
discuss how we apply our algorithm to these simulations to test our scaling
procedures.

\begin{small}
\begin{table}
\begin{center}
\begin{tabular}{rrrrrrrrr}
\hline
\hline
&  $m_{\rm dm}$         & Box      & $\Omega_{\rm m} $ & $\Omega_{\rm b}$ & h            & $\sigma_8$  \\
&  [$h^{-1}~$M$_\odot$] & [Mpc/h]  &            &            &   &             \\
\hline
\tt{W1} & $3.34\times10^{11}$ & $865.16$ & 0.250  & 0.045 & 0.73  & 0.9  \\
\tt{hW1} & $3.34\times10^{8}$ & $86.516$ & 0.250  & 0.045 & 0.73  & 0.9  \\
\tt{W3} & $4.92\times10^{11}$ & $1000.0$ & 0.238 & 0.0416 & 0.732 & 0.761 \\
\tt{hW3} & $4.92\times10^{8}$ & $100.0$ & 0.238 & 0.0416 & 0.732 & 0.761 \\
\hline
\end{tabular}
\end{center}
\caption{The values of some of the basic parameters used in the simulations.
The columns are as follows: (1) The name of the simulation.  (2) The mass of a
dark matter particle. (3) the side of the computational box (4)
The matter density. (5) The
baryon density. (6) The Hubble parameter in units of 100 kms/s/Mpc. (7)
The linear fluctuation amplitude at $z=0$, $\sigma_8$. In both
cases, the primordial spectral index is $n_s = 1$ and the dark energy is
assumed to be a cosmological constant.
}
\label{tab:params}
\end{table} 
\end{small}

\subsection{Direct N-body Simulations}

In this paper we compare results from cosmological simulations at two very
different resolutions: large-volume simulations are used to  study the
performance of our algorithm in reproducing large-scale structure; smaller
volume but higher resolution simulations are used to study its performance in
reproducing the properties of individual dark matter haloes. The latter
simulations are also used as input to a semi-analytic model for galaxy
formation which allow us to compare the properties of the galaxies predicted in
our scaled and true simulations. 

The first of our two cosmologies is that of the Millennium Simulation
\citep{Springel2005a} and is consistent with an analysis of the 1st year of
data from the WMAP satellite \citep{Spergel2003}. The corresponding large and
small volume simulation are labelled as {\tt W1} and {\tt hW1}, respectively.
The second cosmology corresponds to an analysis of the 3rd year WMAP data
\citep{Spergel2007, Sanchez2006}, and we refer to the large and small volume
simulations as {\tt W3} and {\tt hW3}, respectively.  The specific values of
the parameters for the two models can be found in Table 1.  

Each of our four simulations contains $512^3$ dark matter particles. For the
{\tt W3} simulation the size of the box is $L'=1000\Mpc$ and the redshifts at
which we store the particle data are $z'=(0, 0.5, 1.0, 2.0)$.  For {\tt W1} the
corresponding values are $L=865.164\Mpc$ and $z=(0.57, 1.09, 1.68, 2.92)$.
These figures were set so that at each output the respective box sizes and
linear amplitude of fluctuations minimize Eq.~\ref{eq:rms} as described in \S
2.1 for $[R_1, R_2]=[1,10]\Mpc$.  The {\tt hW1} and {\tt hW3} simulations have
the same cosmological parameters as {\tt W1} and {\tt W3}, but have a 10 times
smaller box and 1000 times smaller particle masses. 

The starting redshifts are $z=60.0$ and $z=78.08$ for the {\tt W1} and {\tt W3}
simulations, respectively. The Plummer-equivalent softening length has the same
value in units of the respective box sizes, $\epsilon=L/10000$.  The linear
power spectrum is calculated for each cosmology using the {\tt CAMB} package
\citep{Lewis2000}. The initial density fields are generated using the
Zel'dovich approximation to perturb a set of particles initially distributed on
a cubic grid. Finally, all simulations were evolved with a memory efficient
version of the {\tt Gadget2} code \citep{Springel2005b}.

We note that the random number sequence used to set initial perturbation
amplitudes and phases was the same in the two different cosmologies simulated
at each resolution. This allows us to study subtle differences in the final
halo and matter distributions and provides stringent and clean tests for our
algorithm. 

\subsection{Scaled Simulations: the algorithm in action}

\begin{figure} 
\includegraphics[width=8.5cm]{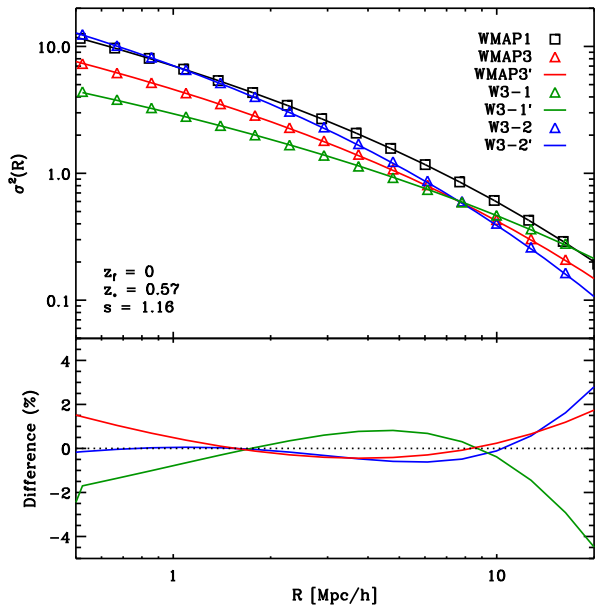} 
\caption{The linear mass variance $\sigma^2(M)$ smoothed with a real-space
top-hat filter as a function of the smoothing radius for various cosmologies.
Symbols in the upper panel show the expected values for our standard WMAP1
(black squares) and WMAP3 (red triangles) cosmologies, and for two cosmologies
where all WMAP3 parameters are retained except that $\Omega_m$ is raised to 0.4
(blue triangles) and reduced to 0.1 (green triangles).  The continuous lines
{\it all} derive from the WMAP1 cosmology but in the red, green and blue cases
our scaling in length and our reassignments of time have been applied in order
to match the corresponding ``WMAP3'' model as well as possible over the range
$1 \Mpc < R < 10 \Mpc$. The quality of these fits is better appreciated in the
bottom panel which shows the difference between symbols and curves as a
percentage.
\label{fig:sigma}} \end{figure}

\begin{figure} 
\includegraphics[width=8.5cm]{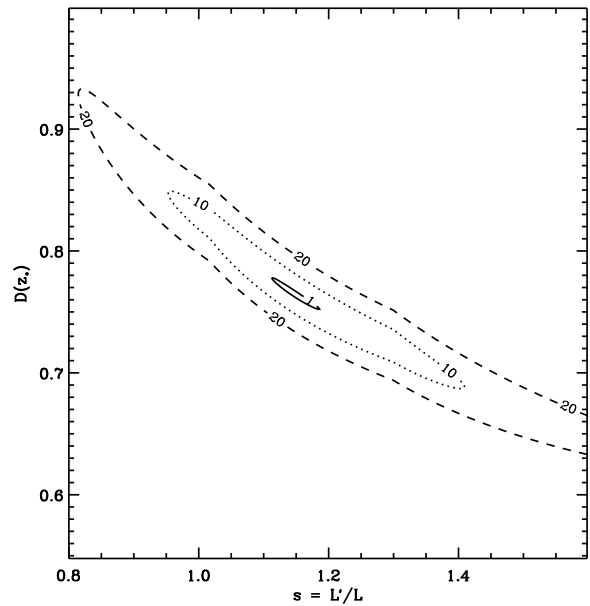} 
\caption{
Constraints on $s$ and $D(z_*)$ obtained by minimizing the logarithmic
difference between the WMAP3 and scaled WMAP1 linear variances over the scale
range corresponding to resolved haloes in the WMAP3 cosmology.  Contours
displayed by solid, dotted and dashed lines indicate the regions in parameter
space that produce rms deviations smaller than $1$, $10$ and $20\%$.  over this
mass rangee (see bottom panel of Fig.~\ref{fig:sigma}). 
\label{fig:contour}} \end{figure}

In this section we give details of how the {\tt hW1} and {\tt W1} simulations
were scaled and perturbed, according to the algorithm described in \S2, to
predict the dark matter clustering and the halo properties expected in the
cosmological model of the {\tt W3} and {\tt hW3} simulations. 

The first step in our procedure is to scale comoving positions and to re-label
redshifts. As discussed in \S2.1, this requires finding a new box-size $L'$ and
redshift $z_*$ for the outputs of {\tt hW1} and {\tt W1} that best reproduce
the linear mass variance at $z=0$ of the target WMAP3 cosmology on resolved
nonlinear mass scales.  

Linear variance is displayed as a function of scale for the original (WMAP1)
and target (WMAP3) cosmologies in the upper panel of Fig.~\ref{fig:sigma},
together with the best fit of the WMAP1 variance to the WMAP3 variance after
scaling lengths and times (see Eq.~\ref{eq:rms}). The range of smoothing radii
used in the fit is $[1-10]\Mpc$ in the WMAP3 cosmology, corresponding to
$[10^{11}-10^{15}]\,\Mass$ for our assumed top-hat filter in real space. This
is the halo mass range expected to host the galaxies observable in future
spectroscopic surveys. 

In the bottom panel of Fig.~\ref{fig:sigma} we can see the quality of the fit.
Over the scales included in the minimization, the mass variance in the scaled
cosmology differs from that in the target cosmology by less than $1\%$.  This
deviation implies a less than $5\%$ difference in the expected number of
``$3\sigma$" haloes between the target simulation and the scaled one, assuming
that the mass function of haloes is well described by the
\cite{PressSchechter1974} model. 

The potential of our algorithm is further illustrated by the blue and red lines
in  Fig.~\ref{fig:sigma}, which show the results of scaling the WMAP1 variance
to match the variance in two more distant cosmologies. $\Omega_m$ is set
to $0.1$ and $0.4$ in the red and blue cases, respectively, with the remaining
parameters kept at their WMAP3 values. For both cases our scaling of WMAP1
reproduces the linear variance of the ``WMAP3'' model  satisfactorily and at a
level comparable to the fit of the real WMAP3 cosmology, although, the scaling
of box size and the relabeling of redshift is more drastic, as expected.  The
values for $s$ are $3.47$ and $0.536$, and for $D(z_*)$ are $0.416$ and $1.311$
for lower and higher $\Omega_m$, respectively.
  
The contour lines in Fig.~\ref{fig:contour} show the constraints on $s$ and on
$D(z_*)$ for our standard WMAP1-WMAP3 scaling. The best fit values are $s=1.156$
and $D(z_*)=0.764$ (corresponding to $z_*=0.57$). The solid, dotted and dashed
lines enclose regions in this two-dimensional parameter space which give rms
differences below $1$, $10$, and $20\%$ respectively,  between the scaled WMAP1
variance and the WMAP3 variance over the range $[R_1, R_2]$ (see bottom panel
of Fig.~\ref{fig:sigma}). 

Fig.~\ref{fig:contour} illustrates the degeneracy between these two parameters.
The banana-like shape of the contours is, in fact, useful because it gives some
latitude in choosing which $s_*$, in the original cosmology should correspond
to $z=0$ in the target cosmology, allowing one to choose $z_*$ to correspond to
one of the stored outputs.

For the second part of our algorithm we have produced low-pass-filtered
displacement and velocity fields for the WMAP1 and WMAP3 cosmologies on a
$512^3$ grid as described in \S2.2. As discussed previously, at each redshift
we included only the contribution of linear modes with $k < R_{nl}^{-1}$. For
$z=0$ this criterion implies $k < 0.23\,\hMpc$, while at $z=2$ it implies $k <
1.01\,\hMpc$. This corresponds approximately to smoothing the displacement fields 
with top-hat filters of radii $4.3\,\Mpc$ and $0.98\,\Mpc$, respectively. 


For the specific simulations we consider in this paper, our results imply that
if we expand the $z=0.57$ outputs of {\tt W1} and {\tt hW1} by a factor of
$1.156$, and then apply the computed large-scale corrections, the results
should be almost identical to the {\tt W3} and {\tt hW3} simulations at $z=0$.
Below we refer to these resulting scaled catalogues as {\tt W3}$'$ and {\tt hW3}$'$
and we examine in detail how close they are to {\tt W3} and {\tt hW3},
respectively.   

\section{COMPARISON WITH DIRECT N-BODY SIMULATIONS}

In this section we assess the accuracy of our procedure by comparing the
properties of dark matter, halo and galaxy catalogues derived from the scaled
{\tt W3}$'$ and {\tt hW3}$'$ simulations with similar catalogues derived from the
true {\tt W3} and {\tt hW3} simulations. 

\begin{figure} 
\includegraphics[width=8.5cm]{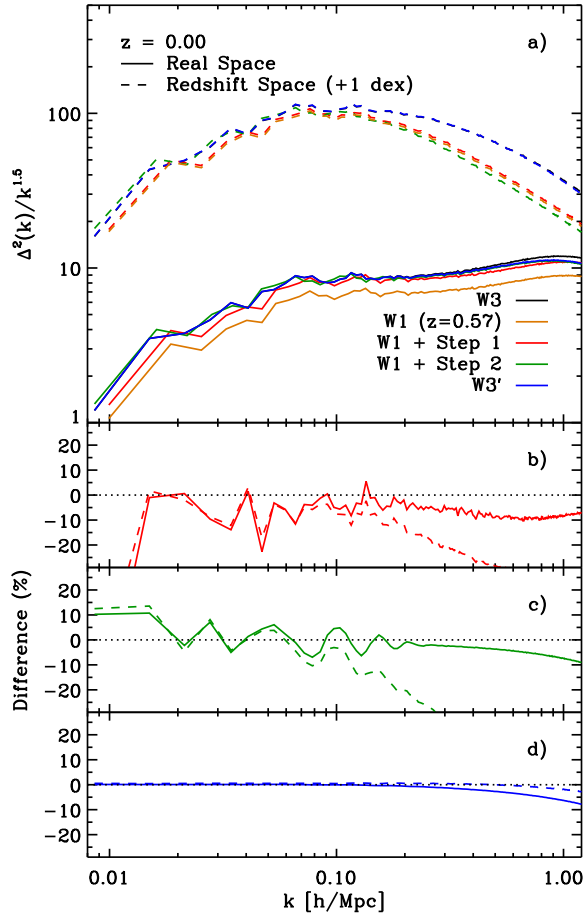} 
\caption{
Comparison of the dimensionless power spectrum, $\Delta^2(k)=k^3 P(k)/2\pi^2$,
measured in the {\tt W3} simulation with those of the {\tt W1} simulation after
different types of scaling.  Note that in panel (a) the redshift-space spectra
have been offset by 1 dex for clarity. 
Poisson shot-noise level.  Black and brown lines show the power spectra of the
dark matter in the {\tt W3} and {\tt W1} simulations, respectively (note that
we show the {\tt W1} simulation at $z=0.57$ and that the black dashed line lies
behind the dashed blue line). Red lines show the result of scaling up the
amplitude of the {\tt W1} simulation at $z=0.57$ to match the linear amplitude
of perturbations in spheres of $8\Mpc$. Green lines display the spectrum after
scaling both the simulation box and the redshift. In both cases, velocities
have been scaled according to Eq.~\ref{eq:linvel}. Finally, blue lines
illustrate the result of including our linear theory correction in addition to
the length and redshift scalings. The deviations after applying the different
procedures can be better appreciated in panels (b), (c) and (d), which show the
fractional difference $\Delta^2_{\tt W1}(k)/\Delta^2_{\tt W3}(k)-1$. In each of
the four panels, power spectra computed in real and redshift space are
displayed as solid and dashed lines respectively. 
\label{fig:pk_dm_alone}} \end{figure}

\subsection{Mass clustering}

\begin{figure} 
\includegraphics[width=8.5cm]{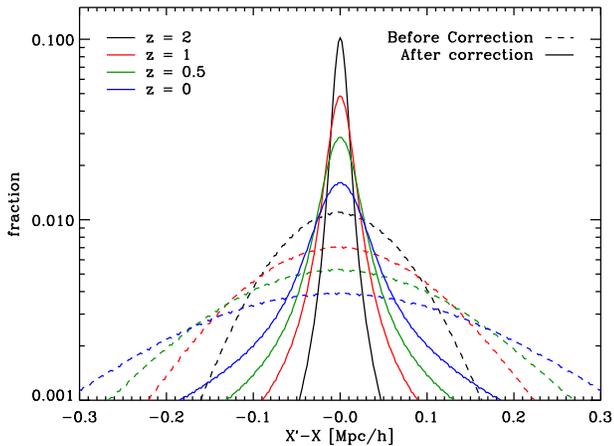} 
\caption{
The distribution of the differences in x-coordinate for dark matter particles
before (dashed lines) and after (solid lines) our large-scale correction is
applied. Different colours show results at different redshifts. 
\label{fig:rmsdm}} \end{figure}

\begin{figure} 
\includegraphics[width=8.5cm]{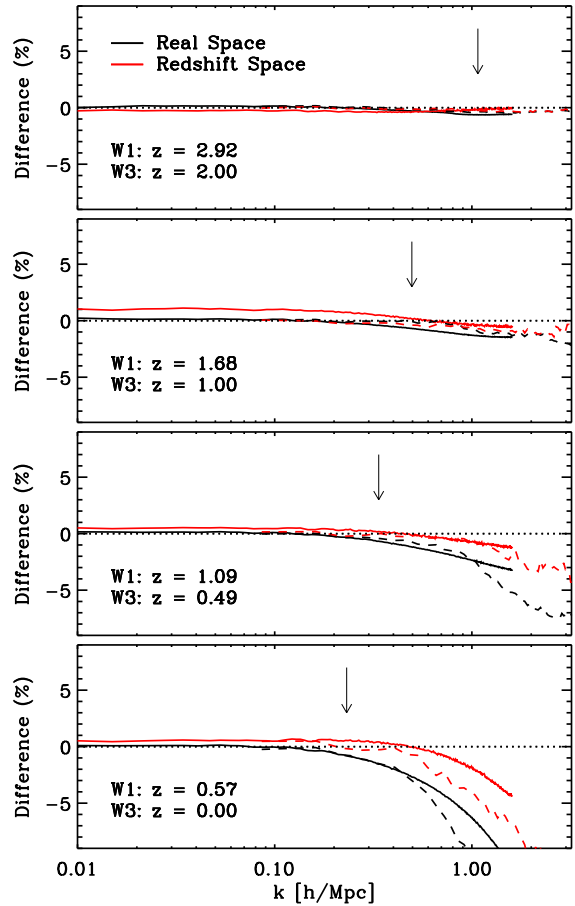} 
\caption{
Relative difference, $\Delta^2_{\tt W3'}(k)/\Delta^2_{\tt W3}(k)-1$, between
the dimensionless dark matter power spectra measured in the scaled and true
simulations as a function of redshift. Solid lines compare {\tt W3}$'$ and {\tt
W3} and dashed lines {\tt hW3}$'$ and {\tt hW3}. Black curves refer to
real-space and red to redshift-space power spectrum. In each panel, the arrow
indicates the wavelength, $k_{nl}$, that satisfies $\Delta^2(k_{nl}) = 1$. This
is the largest $k$-value included in our linear theory corrections. The legend
in each panel indicates the redshifts of the original ({\tt W1}) and target
({\tt W3}) simulation outputs.
\label{fig:pk_dm_z}} \end{figure}

\begin{figure*} 
\begin{center}
\includegraphics[width=18cm]{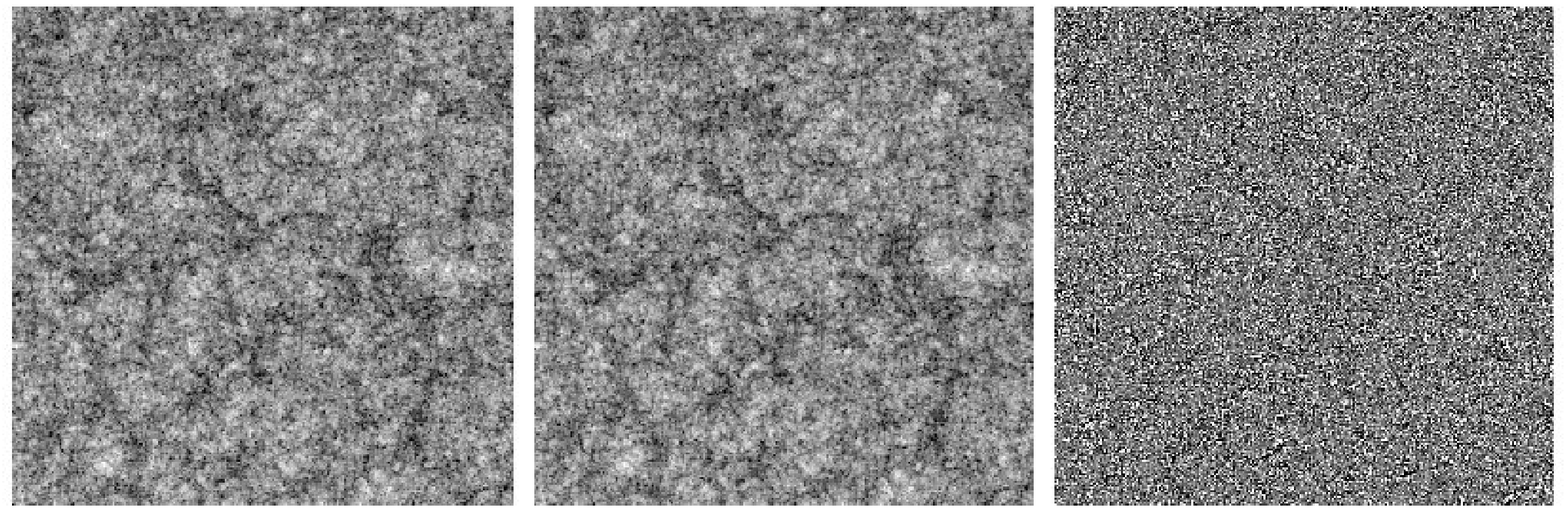} 
\includegraphics[width=18cm]{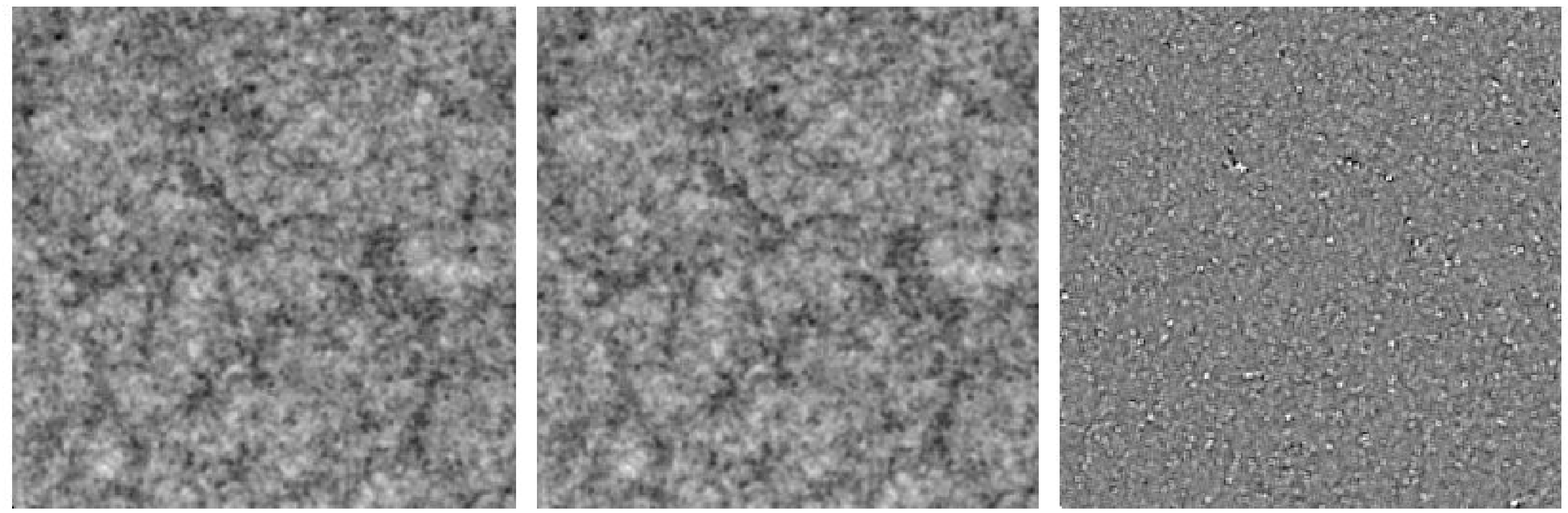} 
\includegraphics[width=18cm]{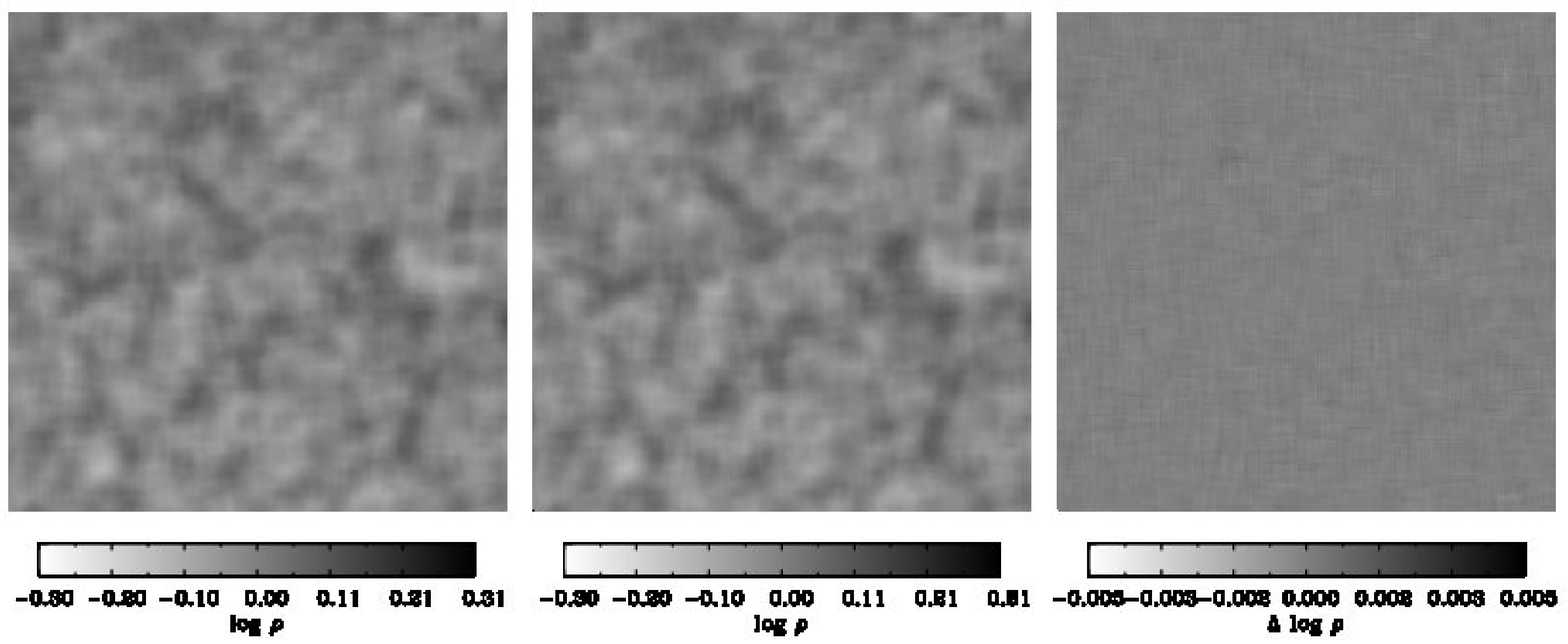} 
\caption{
The mass density field in a slice of thickness $50\Mpc$ and side $1000\Mpc$ at
$z=0$ for the {\tt W3} (left column) and the {\tt W3}$'$ simulations (middle
panel). The difference between these two maps is shown at greatly enhanced
contrast in the right column.  From top to bottom, the dark matter field has
been smoothed with Gaussians of radius $2$, $5$ and $20\,\Mpc$. The intensity
of each pixel is proportional of the logarithm of the density.
\label{fig:pic}} 
\end{center}
\end{figure*}

We start testing our algorithm by examining the spatial distribution of dark
matter particles. Fig.~\ref{fig:pk_dm_alone} shows the dimensionless power
spectra, $\Delta^2 = k^3 P(k)/2 \pi^2$, in real and redshift space, measured in
the {\tt W3} simulation and in the {\tt W1} simulation after three different
scaling procedures have been applied. (The power spectra were computed by
assigning particles onto a $512^3$ grid using the Cloud-in-Cell scheme, and
then Fast Fourier Transforming.)

The first scaling (red lines) consists of adjusting, in a scale-independent
way, the linear amplitude of fluctuations in spheres of $8\Mpc$ of the $z=0.57$
output of {\tt W1} to match the fluctuations of the $z=0$ output of {\tt W3}.
The second scaling (green lines) includes a rescaling of the box size in
addition to the relabeling of the time variable as described in \S3.2. In these
two cases, the velocities were scaled according to Eq.~\ref{eq:linvel}.  The
third scaling (blue lines) uses the linear-theory corrections on large scales
and thus represents our algorithm in full. 

As can be appreciated in the top panel of Fig.~\ref{fig:pk_dm_alone}, all the
described procedures perform moderately well, at least on real space.
However, the closer look given in the lower sub-panels reveals the deficiencies
of each of them.  For the simple amplitude scaling, panel (b) shows that
differences reach $20\%$ even on scales comparable to the fundamental mode of
the simulation box.  As the phases in the two simulations coincide in units of
box size (and {\it not} in units of $\Mpc$), the ratio is very noisy. As shown
in panel (c), these stochastic fluctuations disappear when our length and time
scalings are applied.  Nevertheless, a strong disagreement persists and an
overestimation of nearly $10\%$ remains in the real-space and redshift-space
clustering on large scales. These differences result from the different overall
shape of the linear theory spectra of the {\tt W1} and {\tt W3} simulations,
and from the fact that we have normalised them to agree on nonlinear scales. In
addition, the residuals in panel (c) show an oscillatory pattern. This is a
consequence of the different BAO wavelength in the target and scaled
simulations, a reflection in part of the different $\Omega_{\rm b}$, but mostly
of the rescaling in length.  The last panel (d) demonstrates that these
large-scale features are efficiently removed by our linear correction
procedure.

As can be seen also in panels (b) and (c) of Fig.~\ref{fig:pk_dm_alone}, the
velocity scaling of Eq.~6 does an extremely good job in matching streaming
motions (and thus the redshift-space power spectra) on large scales, but fails
for $k > 0.1 \hMpc$ because it is inappropriate for nonlinear quasi-virialised
motions. Panel (d) shows that our preferred correction on these scales
(Eq.~12) does a much better job.

As we show in Fig.~\ref{fig:rmsdm}, after our large-scale correction the
1D rms difference between the position of particles in the true and ``scaled''
simulations is $0.038\Mpc$ at $z=0$.  This is a factor of $5$ smaller that
the rms difference if only the length and time scalings are
performed ($0.197\Mpc$). At higher redshifts the improvement provided by our
large-scale correction is even greater. For instance, at $z=2$ it reduces the
differences in each dimension by a factor of $10$; the rms values are $0.072$\Mpc
and $0.0073$\Mpc before and after correction, respectively.  

We examine the performance of our full scaling scheme at different redshifts in
Fig.~\ref{fig:pk_dm_z}. Over the range of scales displayed, the dark matter
power spectrum of the {\tt W3} and {\tt W3}$'$ simulations are in very good
agreement. At almost all redshifts the deviations on very large scales
($k<0.1\hMpc$) are of order $0.2\%$ or less.  On smaller scales the amplitude
is systematically lower in the scaled dark matter distribution, a
discrepancy which gets larger at lower redshift. This small difference ($5\%$
in real space and $10\%$ in redshift space at $k \sim 1\hMpc $) is due to
differences in the halo mass-concentration relations.  In the standard picture
for structure formation, the concentration of a halo is related to the mean
density of the Universe at the time of its formation (which is usually defined
in terms of its mass accretion history). Even though our true and
scaled cosmologies have, by construction, the same perturbation amplitudes
at matched redshifts, their growth histories differ when expressed in terms of
redshift. Thus the mean collapse redshift of haloes of given mass will also be
different \citep{Dolag2004,Jennings2009}. As we will see below, this implies
that haloes in the scaled simulations are less concentrated than in the true
simulations. The effect increases with decreasing redshift reaching $10\%$ by
$z=0$.  On the other hand, the absence of wiggles in the residual around the
BAO scale implies that our Zel'dovich-based corrections do appropriately
compensate not only for the linear signal but also, at lowest order, for its
nonlinear degradation.

A visual impression of the quality of the scaling is given by Fig.~\ref{fig:pic}. 
This figure shows the dark matter density field in real space at
$z=0$ measured in $50\Mpc$ slices through the {\tt W3} (left column) and {\tt
W3}$'$ (central column) simulations.  Underdense regions are lighter while
clusters and filaments are darker  The differences between the two maps are
displayed with enhanced contrast in the right column. The density fields have
been smoothed with Gaussian kernels of radius $r_0 = $ $2$, $5$ and $20$ $\Mpc$
(top, middle and bottom rows respectively).  In this way we accentuate the
differences between the dark matter fields on different scales. 

In every row, we can see that the two simulations have identical structures -
voids, filaments and clusters are found at exactly the same locations.  Small
scales, best seen in the upper panel, show a slight positive bias in dense regions
which results from the higher concentration of dark matter haloes in the
true {\tt W3} simulation than in the scaled {\tt W3}$'$ simulation. Small
shifts in the position of clusters appear as black-white dipoles in the right
column.  The discrepancies between the maps are small in all cases and are
strongly reduced when we smooth more strongly.  In the bottom row the two
fields are virtually indistinguishable. This is consistent with our power
spectrum analysis: the matching procedure performs remarkably well on large
scales but does not perfectly reproduce the small-scale nonlinear clustering. 

\subsection{The properties of DM haloes}

\begin{figure} 
\includegraphics[width=8.5cm]{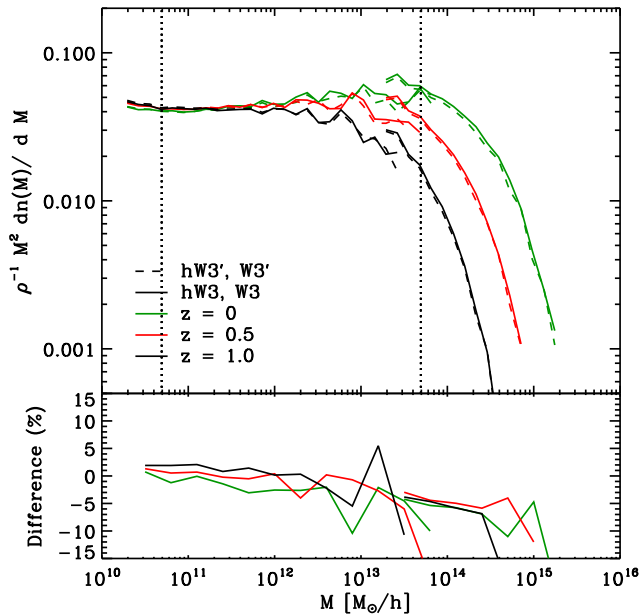} 
\caption{
Comparison of the comoving number density of FoF haloes. We plot the
multiplicity function $\rho^{-1} M^2 {\rm d}N /{\rm d}M$ where $\rho$ is the
mean density of the Universe. The curves, from top to bottom, represent
redshifts $z=0$, $0.5$ and $1.0$.
Solid lines display measurements from the direct simulations {\tt hW3} and {\tt
W3} while dashed lines show results from the scaled catalogues {\tt W3}$'$ and
{\tt hW3}$'$. Vertical dotted lines indicate the 100 particles limit for the
{\tt hW3} and {\tt W3} simulations.  The lower panel shows the difference
``scaled''-true as a percentage.  
\label{fig:massfn}} 
\end{figure}

\begin{figure*} 
\includegraphics[width=17cm]{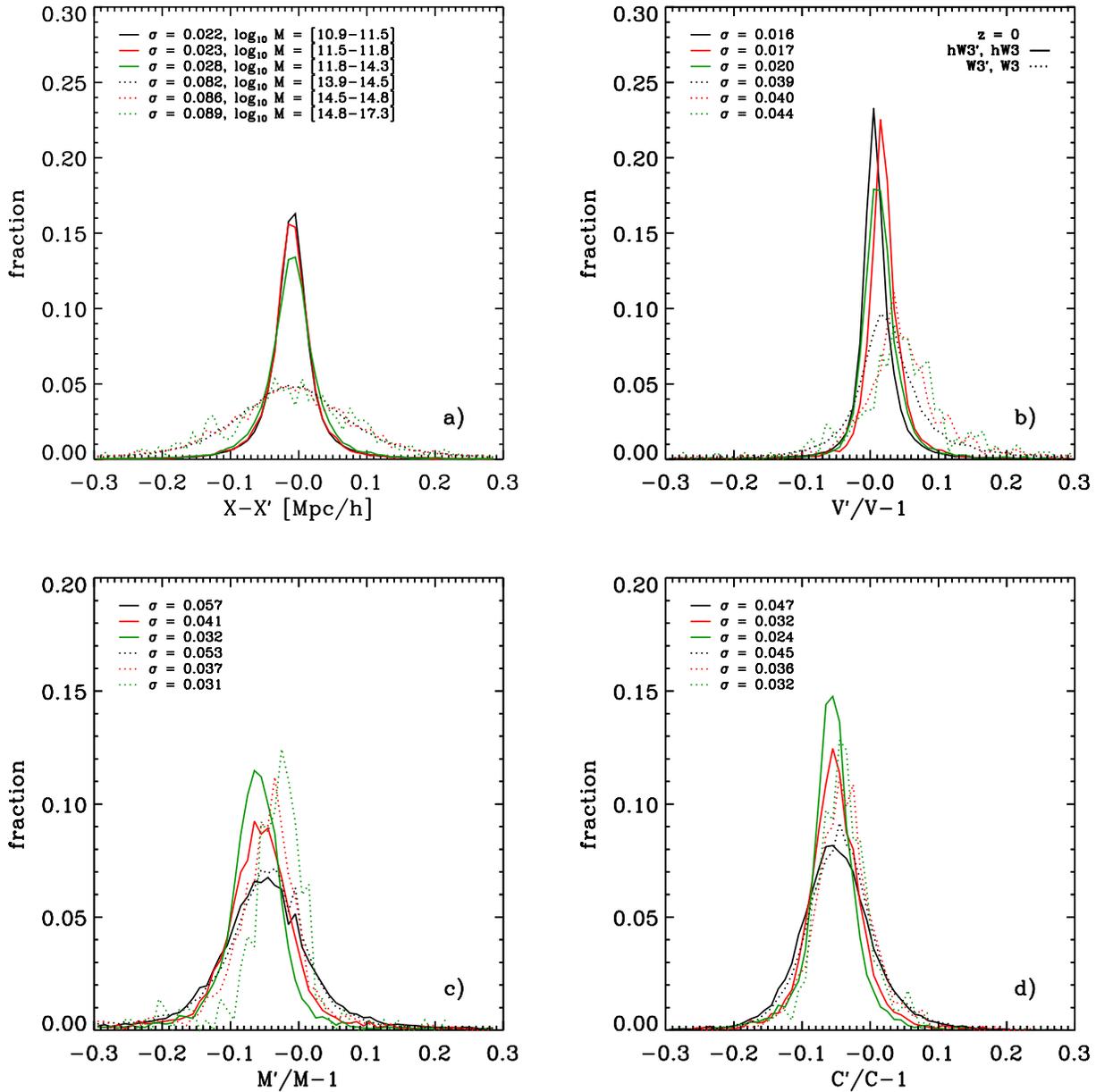} 
\caption{
The distribution of the differences in position, velocity modulus, mass and
concentration for FoF haloes at $z=0$ in {\tt W3} and {\tt W3}$'$ (dotted
lines) and in {\tt hW3} and {\tt hW3}$'$ (solid lines).  Results are shown for
$6$ disjoint mass ranges.  Black, red and green curves show distributions
for haloes with  $[200,700]$, $[700,1500]$ and $>1500$ particles in {\tt W3}
and {\tt hW3}. The standard deviation of the distributions is given in the
legend of each panel.   
\label{fig:dist}} \end{figure*}

\begin{figure*} 
\includegraphics[width=17cm]{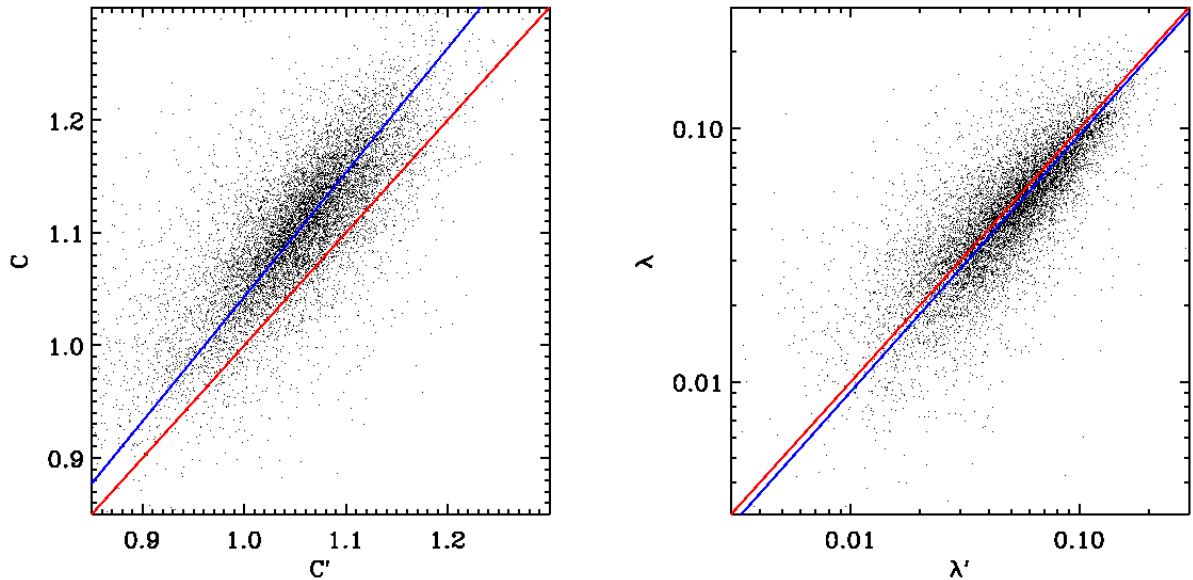} 
\caption{
Comparison of the concentration (left panel) and of the spin parameter (right
panel) for haloes identified in the true (unprimed) and scaled (primed)
simulations at $z=0$. For clarity, we have plotted the result only for haloes
containing more than $500$ particles.  The red lines indicate the result
expected if there were no systematic biases between the halo properties in our
two simulations (i.e. the one-to-one relation), while the blue lines indicate
the straight line that minimises the rms perpendicular residuals.
\label{fig:conspin}} \end{figure*}

\begin{figure} 
\includegraphics[width=8.5cm]{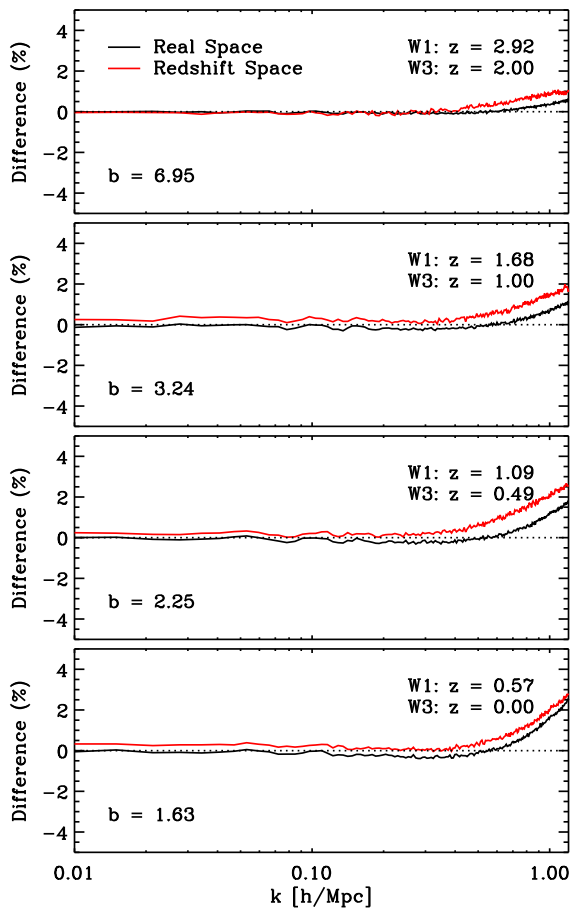} 
\caption{
Relative difference, $\Delta^2_{\tt W3'}(k)/\Delta^2_{\tt W3}(k)-1$, between
the dimensionless halo power spectra measured in the scaled and true
simulations as a function of redshift.  We have used only matched FoF haloes
with at least $50$ particles in {\tt W3}. This limit corresponds to a mass of
$2.6\times10^{12}\Mass$ in the WMAP3 cosmology.  The bias of the sample,
defined as the mean value of $b = \sqrt{ \Delta^2_{\rm h}(k)/\Delta^2_{\rm
dm}(k)}$ on scales $k < 0.1\hMpc$, is shown in the legend of each subpanel
and is seen to vary strongly with redshift. 
\label{fig:pk_h_z}} \end{figure}

\begin{figure} 
\includegraphics[width=8.5cm]{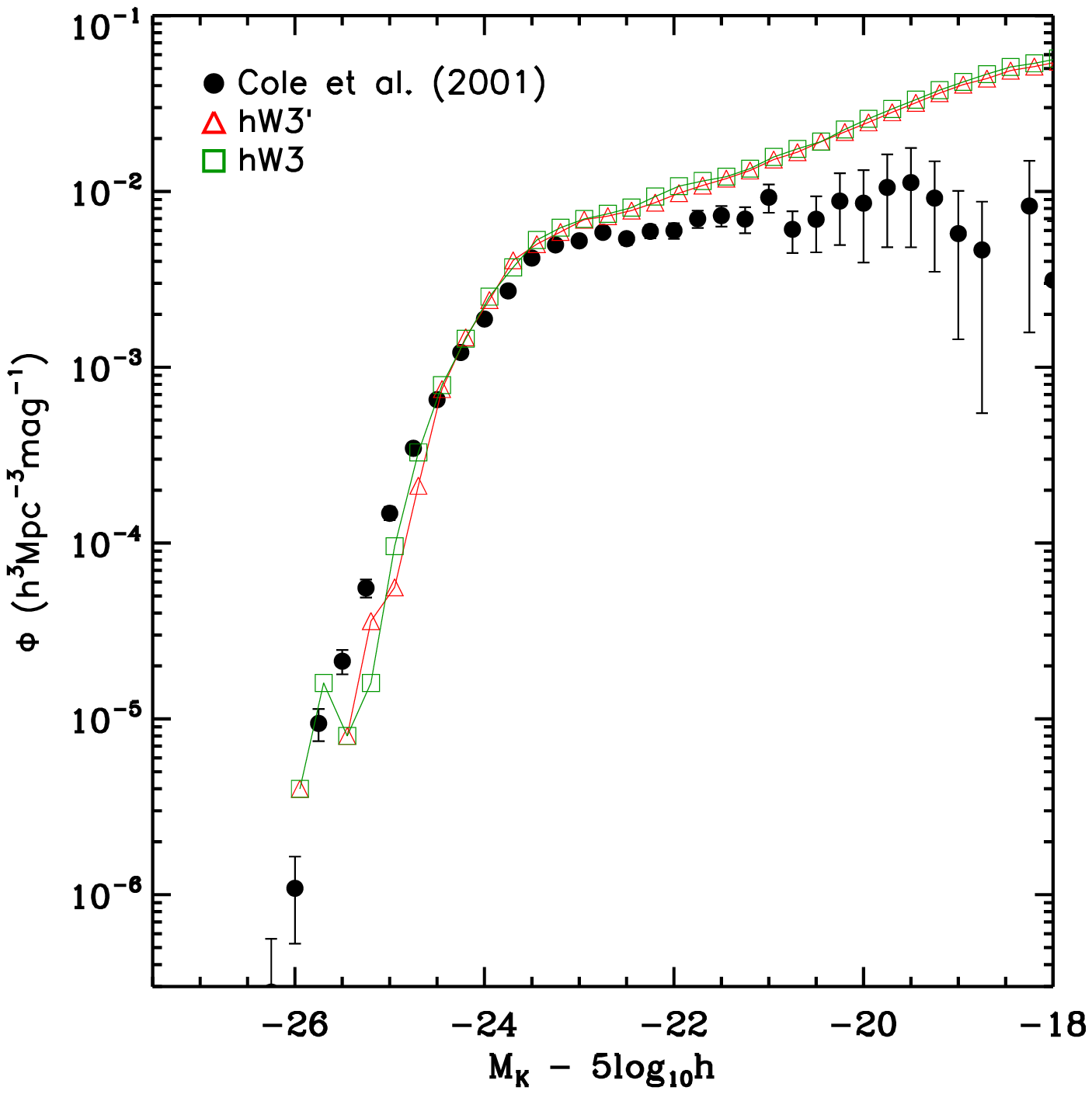} 
\caption{ 
The K-band luminosity function at $z=0$ constructed from the {\tt hW3} (green
squares) and {\tt hW3}$'$ simulations (red triangles).  For comparison, we have
also included the luminosity function of the 2dFGRS galaxies as computed from
Cole~et al.(2001). 
\label{fig:lf}} \end{figure}

We now compare the distribution of dark matter haloes in the true ({\tt W3})
and scaled {\tt W3}$'$ simulations. These catalogues were built using a FoF
group finder \citep{Davis1985} with a linking length equal to $20\%$ of the
mean interparticle separation, which approximately selects objects with $200$
times the {\it mean} density. At $z=0$ the number of haloes with at least $50$
particles in the {\tt W3} and {\tt W3}$'$ catalogues is $138,284$ and
$131,356$, respectively. These figures are $109,651$ and $109,176$ for the {\tt
hW3} and {\tt hW3}$'$ catalogues. 

Fig.~\ref{fig:massfn} shows the abundance of dark matter haloes as a function
of mass in our two simulations. Over the mass range covered by the catalogues,
the mass functions are in very good agreement. Differences are below $10\%$
over $4$ orders of magnitude in mass and over $7$ orders of magnitude in number
density! However, focusing on the bottom panel, where differences between the
curves can be seen in more detail, it is evident that there are significant
systematic discrepancies. 

The scaled simulations {\tt W3}$'$ and  {\tt hW3}$'$ seem to underpredict the
abundance of haloes compared to the {\tt W3} and  {\tt hW3} simulations, or
equivalently, FoF masses are systematically higher in the true simulations than
in the scaled simulations. In particular, the difference in abundance seems
larger for high-mass haloes, this explains the fact that the total number of
haloes differs by $5\%$ between {\tt W3} and {\tt W3}$'$, but only by $0.5\%$
between {\tt hW3} and {\tt hW3}$'$.  There are a number of possible reasons for
this disagreement. A $\sim5\%$ disagreement is expected due to the residuals in
the mass variance fitting (See Fig.~\ref{fig:sigma}). In addition, a fixed
linking length identifies objects with lower mean overdensities relative to the
mean density if their concentrations are lower, therefore underestimating the
mass of objects identified at low redshift.  Finally, an important assumption
of our work is that the halo mass function is universal (i.e. independent of
redshift, cosmological parameters, etc) when the masses are expressed in terms
of $\delta_c/\sigma(M)$, as predicted by Press-Schechter theory.  However,
recent simulation results suggest that this picture may work no better than
$10\%$, and that properties of the density field other than the shape of the
linear variance as a function of smoothing radius may be important
\citep{Tinker2008, Manera2009}. 

In order to provide a more detailed comparison between our catalogues we have
matched individual haloes in {\tt W3}$'$ and {\tt W3}. We do this by selecting
all the DM particles that belong to each halo with more than $50$ particles in
{\tt W3} and then finding the halo that contains most of these particles in
{\tt W3}$'$.  Due to statistical fluctuations not every halo in the {\tt W3}
({\tt hW3}) catalogue can be matched  to one in the {\tt W3}$'$ ({\tt hW3}$'$)
catalogue.  Nevertheless, of the $138,284$ ($109,651$) haloes with more than
$50$ particles in {\tt W3}  ({\tt hW3}), only $16$ ($21$) are unmatched to a
halo with at least $20$ particles in {\tt W3}$'$ ({\tt hW3}$'$).

The object-by-object comparison that results from this matching is presented in
Fig.~\ref{fig:dist}. In this plot we present the distribution of differences in
the position, velocity, mass and concentration of matched FoF haloes for the
{\tt W3}-{\tt W3}$'$ (dotted lines) and for the {\tt hW3}-{\tt hW3}$'$
catalogues (solid lines).  We show the results at $z=0$ and for haloes in $6$
different mass ranges, but we have checked that our results behave in a similar
way at different redshifts, and for other mass ranges. 

In the top left panel we can see that the centre of mass positions of the FoF
haloes, agree remarkably well.  For each of the 3 spatial coordinates,
discrepancies are typically $85$\kpc between the $1\Gpc$ simulation, but only
$30$\kpc between the $100\Mpc$ simulations.  These uncertainties are less than
half the typical virial radii of the haloes.  This agrees with the small
differences seen in Fig.\ref{fig:pic}. We note that these differences increase
roughly by a factor of two if our large-scale correction is not included.  In
panel (b) we display differences in the 1D velocity modulus ($V^2=V_{x}^2 +
V_{y}^2 + V_{z}^2$) of the centre of mass of our haloes.  We find that this
quantity is also matched very accurately: better than $\sim5\%$ rms, and with
no measurable bias for both pairs of simulations in all the mass bins
inspected.

On the other hand, the FoF masses are slightly less precise.  The uncertainty
in our scaled halo mass is just under $10\%$ with a systematic underestimation
of $\sim 5\%$. We also note that the accuracy at which we match this is greater
for our high-resolution simulations. 

In contrast, in panel (d), we compare concentrations of FoF haloes using
the ratio $C=V_{\rm max}/V_{200}$, where $V_{\rm max}$ is the maximum circular
velocity of the main subhalo of the FoF halo and $V_{200}$ is the circular
velocity at $R_{200}$, the radius enclosing a mean density 200 times the
critical value.  The discrepancies here are typically larger in the high-resolution
pair, although still under 10\% with a bias of about $5\%$. The lower mean
concentration of haloes in {\tt W3}$'$ is responsible for the underestimation
of the real-space power spectra on small scales seen in Fig.~\ref{fig:pk_dm_z}.
We expect this to impact the spatial distribution of galaxies, if at all, only
on very small scales.

In Fig.~\ref{fig:conspin} we show the accuracy with which our scaling procedure
is able to reproduce two important properties of dark matter haloes. In the
left panel we compare concentrations (as defined above) of haloes identified in
the scaled ($C'$) and true simulations ($C$). In agreement with
Fig.~\ref{fig:dist}, there is a small bias in the average value which results
from differing halo growth with redshift in our two simulations: the
concentration of a halo is underestimated by $0.04$ in the scaled simulations.
In the right panel we compare the spin parameter defined as $\lambda =
\vect{J}/(\sqrt{2} M_{200} V_{200} R_{200})$ where $\vect{J}$ is the halo
angular momentum and $M_{200}$ is the total mass (both within $R_{200}$)
\citep{Bullock2001}. This panel demonstrates that, albeit with scatter, spins
are well reproduced by our velocity and position scaling.  
  
We now explore whether the differences displayed in Fig.~\ref{fig:dist} and
Fig.~\ref{fig:conspin} produce a noticeable effect on halo
clustering by comparing the power spectra of the matched catalogues.
Fig.~\ref{fig:pk_h_z} presents results in real space (black lines) and in redshift
space (red lines) at $z=0$, $0.5$, $1$ and $2$. As found above for the dark
matter catalogues, on large scales the two halo power spectra are almost
identical. In real space the differences are less than $0.1\%$ while in
redshift space they are still below the percent level. This behaviour repeats
at all redshifts. On smaller scales, haloes in {\tt W3}$'$ tend to be more
clustered than their counterparts in {\tt W3}.  Overall, the distributions
agree, over all relevant scales, within $1\%$, an uncertainty much smaller than
those likely to be introduced by our poor knowledge  of the galaxy formation
process.    

It is important to note that in these measurements we have used haloes above a
fixed mass. As a result, the samples displayed at different redshifts
correspond to peaks of different height in the density field, as illustrated by
the bias values given in each panel of Fig.~10. Thus, the fact that we recover
the expected clustering at all redshifts suggests that the bias-mass
relationship is very well reproduced by our algorithm, even for very rare
peaks. 

\begin{figure} 
\includegraphics[width=8.5cm]{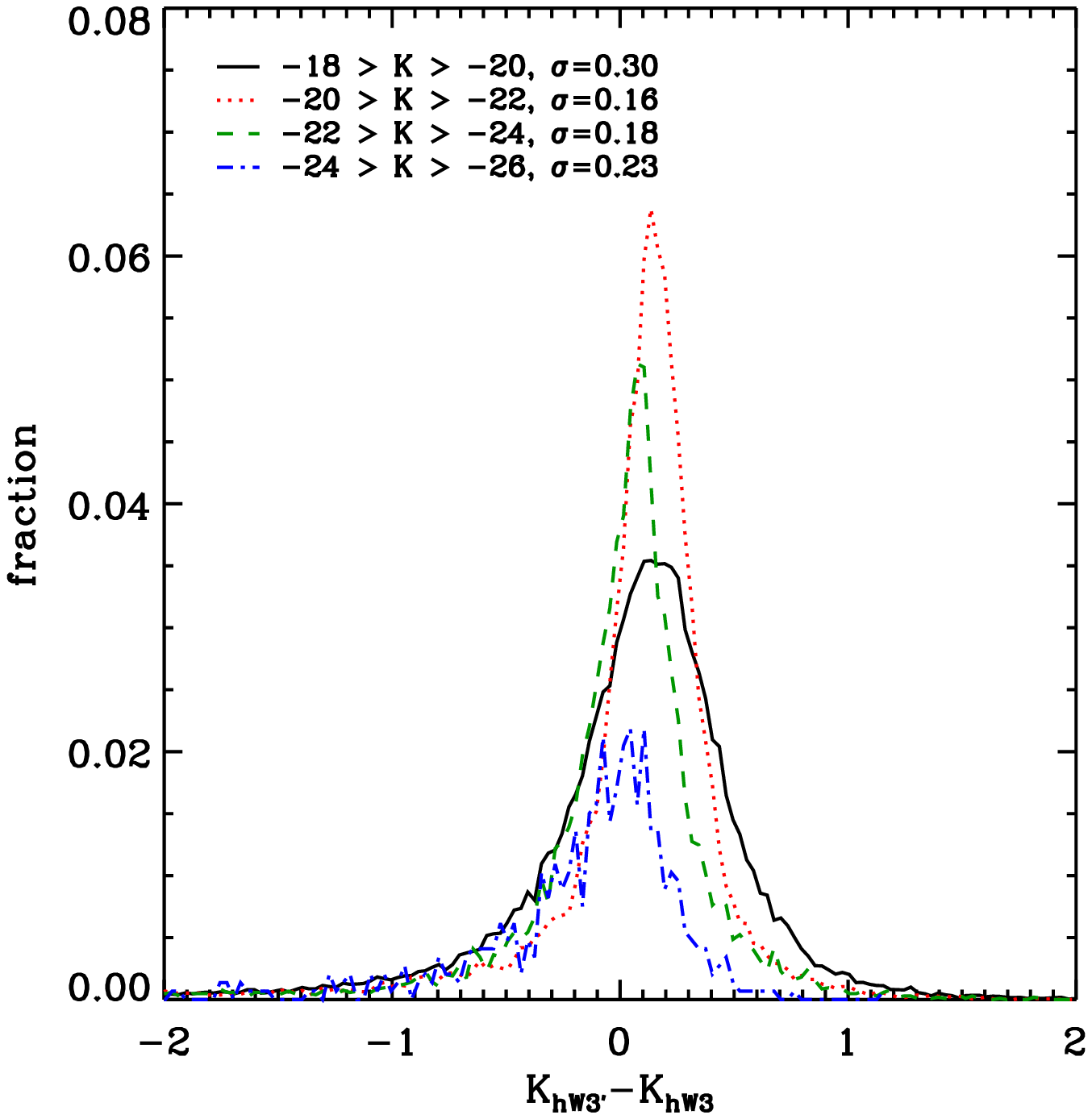} 
\caption{The differences in K-band absolute magnitude for central galaxies in
the {\tt hW3} and {\tt hW3}$'$ catalogues. These galaxies represent the
$\sim56\%$ of all the galaxies with $M_{K} - 5 \log_{10} h < -18$ in our
semi-analytic catalogues. The median of the distribution for the sample as a
whole is $0.135$ while the dispersion is $0.233$.
\label{fig:kband}} \end{figure}

\subsection{Models for galaxy formation}

The final step in the creation of galaxy mock catalogues is to connect the
predictions for dark matter haloes with the observable properties of the
galaxies which they host. For this, a crucial requirement is the construction
of accurate merger histories.  Thus we now explore how well our algorithm
performs in this respect \citep[see][for an attempt to describe merger
histories in a cosmology-independent way]{Neistein2009}. We do this by
comparing the galaxies built using {\tt hW3}$'$ with those built using an
identical procedure on {\tt hW3}.

There are several different approaches to create realistic galaxy populations
in the dark matter haloes of N-body simulations. Here we use so-called
semi-analytic models of galaxy formation in which the key physical processes
that determine the properties of galaxies are followed from very early times in
a self-consistent way within the hierarchically growing halo population.  Such
semi-analytic modelling represents a hard test for our algorithm since the
final properties of galaxies depend not only on the present-day mass of their
host haloes but also on their {\it whole} assembly histories -- merger rates,
progenitor mass function, formation times, spins -- and indirectly on the
properties of the environment that affect these quantities.   

The specific model we use in this paper is that developed by the Munich group
as described in \cite{DeLucia07}, with the parameters of \cite{Wang08}. This
model roughly reproduces the observed $z=0$ K-band luminosity function (among
other galaxy properties) in the WMAP3 cosmology, and is adequate for our tests.
For both {\tt hW3} and {\tt hW3}$'$ we have generated halo merger trees with
the same time resolution as in the Millennium Simulation, i.e. roughly $300$
Myr for $z<1$.  All levels in the two merger trees are set to correspond
appropriately after scaling lengths and masses and reassigning redshifts.  This
involves matching outputs in the {\tt hW3} and the original {\tt hW1}
simulations as above. 

The present-day K-band luminosity functions computed by this model are shown in
Fig.\ref{fig:lf}. The green triangles refer to galaxies in the true model {\tt hW3},
while the blue squares refer to galaxies in the scaled model  {\tt hW3}$'$.
For comparison we have also included the observed luminosity function from
\cite{Cole01}. In both cases our galaxy samples are complete down to $M_{K} -
5\log h = -18$.  Above this magnitude threshold the catalogues contain $76562$
and $71652$ galaxies for {\tt hW3} and {\tt hW3}$'$, respectively. Overall the
agreement is quite impressive, especially if one remembers that we are
comparing results for simulations run with significantly different sets of
cosmological parameters.  However, discrepancies exist: there is a small
underestimate of the number of galaxies brighter than a given magnitude in the
{\tt hW3}$'$ catalogue. This difference is weakly magnitude-dependent and
reaches $5\%$ at the faintest magnitudes. It is consistent with the differences
in the halo mass function seen in Fig.~\ref{fig:massfn}.

A further comparison of the results of the semi-analytic modelling is presented
in Fig.~\ref{fig:kband}. In this plot we show the distribution of K-band
magnitude differences, $K_{\tt hW3}-K_{\tt hW3'}$, for the central galaxies of
``matched" haloes (c.f. \S4.2). The predicted properties of galaxies in the
WMAP3 cosmology are quite accurately reproduced by our scaled simulations.
Typical random differences are below $0.25$ magnitudes over the whole magnitude
range and systematic errors are below $0.13$ magnitudes.

\section{Conclusions}

In this paper we have introduced an algorithm that allows us to scale the
output of a cosmological N-body simulation carried out for one specific set of
cosmological parameters (for example the WMAP1 parameter set of the Millennium
Simulation) so that it faithfully represents the growth of structure in a
different cosmology (for example, one with the lower fluctuation amplitude
preferred by analyses of the WMAP3 data).  Our method builds on the fact that
the {\it linear} fluctuation amplitude as a function of smoothing scale is
sufficient to determine not only the mass function and assembly history of
haloes (according to extended Press-Schechter theory) but also the ensemble
properties of the large scale structure (in the Zel'dovich approximation). We
therefore re-scale the box-size of the simulation and reassign redshifts to the
outputs so that the shape of the smoothed linear power spectrum matches that of
the target cosmology on the scales which produce resolved dark matter haloes.
This scaling and redshift reassignment then imply a rescaling both of particle
mass, and of the nonlinear ``thermal'' velocities in each output. Finally, we
use the Zel'dovich approximation to modify the amplitude of the long-wavelength
quasilinear modes in order to account for the difference in shape on large
scales between the smoothed linear power spectrum of the original cosmology
(after scaling) and that of the target cosmology.  This modification is
effectively a rescaling of the displacements and peculiar velocities produced
by each individual long-wavelength (low $\vect{k}$) mode.

We tested our method by applying it to an N-body simulation based on the WMAP1
parameters to mimic the mass distribution expected in the WMAP3
cosmology. We compared the results with those from a matched N-body simulation
run directly in the target cosmology (WMAP3). These two cosmological models
span the range of power spectrum shape and amplitude among the models currently
considered plausible. Thus we expect our comparison to give conservative
estimates of the accuracy of our method. 
  
We found that on large scales the target mass distribution is recovered to high
precision ($<1\%$ on scales $k<0.2\,\hMpc$) in both real and redshift space.
On smaller scales the differences increase (~$5\%$) and can be attributed to
differences in the halo mass-concentration relation between the true
and scaled simulations. These differences are smaller when one considers the
clustering of dark matter haloes, -- the distribution actually relevant for
galaxy clustering. In the latter case, differences are at the subpercent level
over the full range of scales we examined. We also tested the algorithm on an
object-by-object basis concluding that the position of individual FoF haloes is
reproduced to better than $90\,\kpc$, and their velocities and FoF masses to
better than $10\%$. 

The dispersion in the matched properties introduces no appreciable effects in
the overall clustering statistics of the halo population. Finally for a
representative semi-analytic galaxy formation model, the K magnitudes of
central galaxies are accurate to $0.25$ magnitudes with a systematic error of
$0.13$ magnitudes.

Understanding the connection between galaxy formation and the background
cosmological model is an important challenge that cosmologists will face in the
near future. A key step towards this is the creation of realistic physically
based theoretical models for the galaxy distribution for a wide range of
cosmologies. Our method offers a route to this goal.  Its relatively small
computational cost enables exploration of a multidimensional cosmological
parameter space using a suitable N-body simulation of a single cosmology. In
this way the task of creating realistic mock galaxy catalogues for all allowed
cosmological parameter sets is greatly simplified.  It also becomes possible to
update existing high-resolution simulations to any favoured cosmological model
at a small fraction of the original computational cost.  These goals are
important not only to understand the physical processes underlying galaxy
formation, but also to ensure optimal exploitation of the information encoded
in the very large (and expensive) survey datasets that are currently being
assembled. Our scheme makes it possible to study how uncertainties in galaxy
formation physics will affect our ability to constrain cosmological parameters
through analysis of next-generation galaxy surveys. 

\section*{Acknowlegdments}

We would like to thank C. Baugh, S. Cole and M. Schneider for useful comments and a
careful reading of this manuscript.

\bibliographystyle{mn2e} \bibliography{scl}

\label{lastpage} \end{document}